\documentclass{edbk} 
\usepackage[dvips]{epsfig}
\usepackage{edbkps}

\kluwerbib

\def\be{\begin{equation}}
\def\ee{\end{equation}}

\def\msun{M_{\odot}}

\def\Ti{T_{{\rm out},i}}
\def\Te{T_{{\rm out},e}}

\begin{document}

\articletitle[Accretion Disk]{A New Parameter In Accretion Disk Model}

\author{Feng Yuan}

\affil{Department of Astronomy, Nanjing University, Nanjing 210093, China}
\email{fyuan@nju.edu.cn}

\begin{abstract}
Taking optically thin accretion flows as an example, we investigate the 
dynamics and the emergent spectra of accretion flows with different outer 
boundary conditions (OBCs) and 
find that OBC plays an important role in
accretion disk model. This is because the accretion equations describing the
behavior of accretion flows are a set of {\em differential} equations,
therefore, accretion is intrinsically an initial-value 
problem. We argue that optically thick accretion flow should also 
show OBC-dependent behavior. 
The result means that we should seriously consider 
the initial physical state of the accretion flow such as its angular 
momentum and its temperature. An application example to Sgr A$^*$ is presented. 
\end{abstract}

\section{Introduction}
It has long been assuming that the parameters describing the accretion flow
include the accretion rate, the mass of the central black hole,
the viscosity parameter, and the parameter 
describing the strength of the magnetic
field in the accretion flow. Once these parameters 
are given, we can obtain almost all the 
information of the accretion flow including the dynamics and the emergent
spectrum. However, the set of equations describing the accretion flow
are nonlinear differential equations, therefore it is intrinsically 
an initial-value problem. Thus the outer boundary condition (OBC)
possibly plays an important role. 

On the other hand, the complicated astrophysical environments
make the physical states of the accreting gas at 
the outer boundary $r_{\rm out}$,
such as its temperature and angular momentum, various.
For example, in semi-detached binary system, where the critical 
Roche lobe is filled up and the accretion of matter takes 
place through the inner Lagrangian point, the angular momentum of the accreted
gas should be high; while in detached binary system the accretion 
matter is stellar winds therefore their angular momenta are much lower
(Illarionov \& Sunyaev 1975). 
In the nuclei of galaxies, where the supply of the accretion matter 
is  unclear, the initial physical states of the accretion flows 
should be more complicated.
The complexity of astrophysical environments makes it important 
to investigate the role of OBC in accretion disk model.

\section{The role of OBC in optically thin accretion flows}

\begin{figure}
\vskip -0.4cm
\epsfig{file=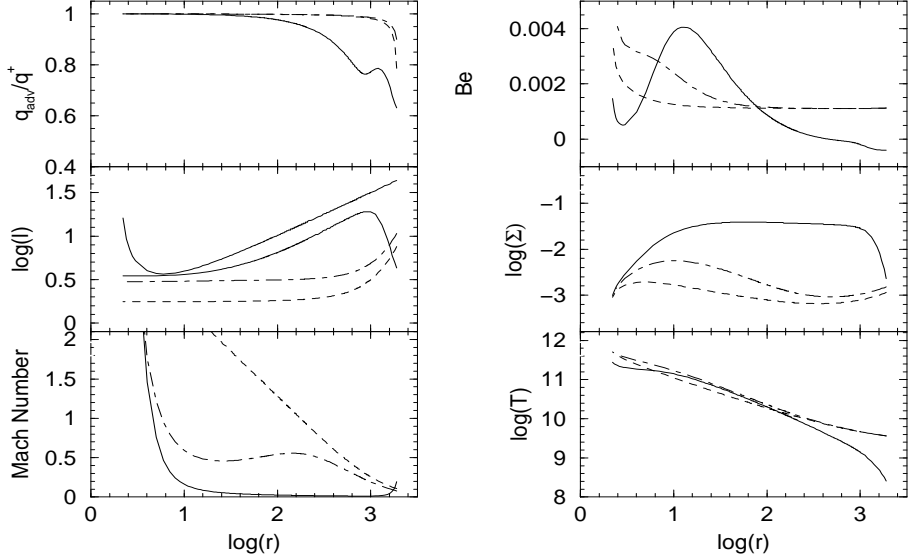, width=7.5cm, height=12.cm,angle=270}
\vskip -0.4cm
\caption{
Solutions for one temperature global solutions with different OBCs for
$ M=10 \msun$, $\dot{M}=
10^{-3} \dot{M}_{\rm E}$ and $\alpha=10^{-2}$. The solid, dot-dashed and
dashed lines represent $(T_{\rm out}, \lambda_{\rm out}(\equiv 
v/c_s \equiv v/\sqrt{p/\rho}))$=($2 \times
10^8 {\rm K}, 0.4), (3.6 \times 10^9 {\rm K}, 0.08)$ and
$(3.6 \times 10^9 {\rm K}, 0.107$) respectively. The units of $\Sigma, T$
are ${\rm g \ cm^{-2}}$ and K, $r$, $Be$ and $l$ are in $c=G=M=1$ units.
Mach number is simply defined as $v/c_s$. The upper-left
plot represents the ratio of the advected energy to the viscous
dissipated energy. Adopted from Yuan (1999).
}\label{fig1}\end{figure}

\begin{figure}
\vskip -0.4cm
\epsfig{file=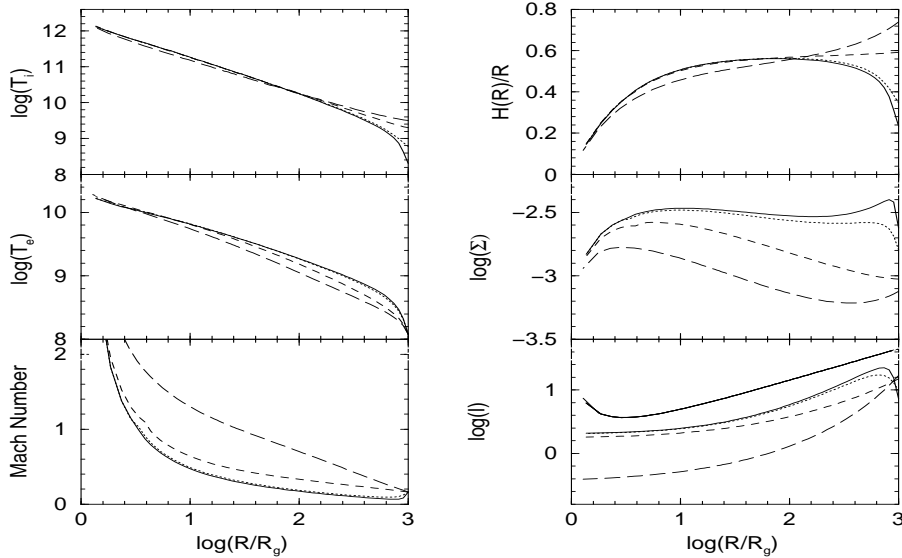, width=7.5cm, height=12.cm,angle=270}
\vskip -0.4cm
\caption{
Solutions for two temperature global solutions with different $\Ti$.
The solid line (type I solution) is for $\Ti=2 \times 10^8K$,
the dotted line (type I) for $\Ti=6 \times 10^8K$,
the dashed line (type II) for $\Ti=
2 \times 10^9K$ and the long-dashed line (type III) for $\Ti=3.2 \times 10^9K$.
Other OBCs are $\Te=1.2 \times 10^8K$ and $\lambda_{\rm out}=0.2$.
The outer boundary is set at $r_{\rm out}=10^3r_{\rm g}$. Other parameters are
$\alpha=0.1, \beta=0.9, M=10^9 \msun$ and $\dot{M}=10^{-4} \dot{M}_{\rm Edd}$.
The units of $\Sigma$ and $T$ are  ${\rm g \ cm^{-2}}$ and K. Adopted
from Yuan et al. (2000).
}\label{fig2}\end{figure}

In previous papers (Yuan 1999; Yuan et al. 2000), taking optically 
thin accretion onto a black hole as an 
example, we calculated the dynamics and the 
emergent spectrum of one- and two-temperature accretion plasma 
by self-consistently solving the radiation hydrodynamical 
equations. For the one temperature case, only bremsstrahlung emission
and its Comptonization are considered, while for the two temperature case,
synchrotron emission and its Comptonization are also included.
We concentrated on the role of OBC by setting the same 
``general parameters'' such as accretion rate, viscosity
parameter and black hole mass while adopting different OBCs.
We adopted the temperature $T_{\rm out}$ and
the ratio of the radial velocity to the local sound speed 
$\lambda_{\rm out}$
(or, equivalently, the angular velocity $\Omega_{\rm out}$) at a certain outer
boundary $r_{\rm out}$ as the outer boundary conditions
and found that in
both cases, the topological structure and the profiles of angular momentum
and surface density of the flow differ greatly under
different OBCs, as shown by Figures 1 (for a one-temperature plasma) 
and 2 (for a two-temperature plasma; only the ions temperature $\Ti$ varies: 
for other cases, see Yuan et al. 2000). 
In terms of the topological structure and the profile
of the angular momentum, three types of solutions are found.
When $T_{\rm out}$ is relatively low,
the solution is of type I. When $T_{\rm out}$ is relatively high and
the angular velocity $\Omega_{\rm out}$ is higher than a critical 
value $\Omega_{\rm crit}$,
the solution is of type II. Both types I and II possess small sonic radii,
but their topological structures and angular momentum profiles
are different. When $T_{\rm out}$ is high but the
angular velocity is lower than $\Omega_{\rm crit}$, the 
solution becomes of type III,
characterized by a much larger sonic radius.
Similar transition has been found previously in the context
of adiabatic (inviscid) accretion
flow by Abramowicz \& Zurek (1981). In that case,
they found that when the specific angular momentum of the flow
decreased across a critical value,
a transition from a disk-like accretion pattern (with small sonic radii) to
a Bondi-like one (with large sonic radii) 
would happen (Abramowicz \& Zurek 1981;
Lu \& Abramowicz 1988). Here in this paper we find that this
transition still exist when the flow becomes viscous, confirming
the prediction of Abramowicz \& Zurek (1981).
Figure 3 shows the emergent spectrum of the solutions presented in 
Figure 2. Considering that they possess the same ``general'' parameters
the discrepancy among the spectra completely caused by 
the difference of OBC is impressive. At last, we should emphasize that 
such ``OBC-dependent'' effect on the spectrum 
has relation with the value of $r_{\rm out}$: 
the smaller $r_{\rm out}$ is, the more significant the effect becomes. Thus, 
this effect should be very obvious in the accretion flow 
where standard thin disk-ADAF transition occurs. As a result, 
some confusing problems can 
be promisingly solved (Yuan \& Yi, in preparation).

\begin{figure}
\vskip -0.cm
\epsfig{file=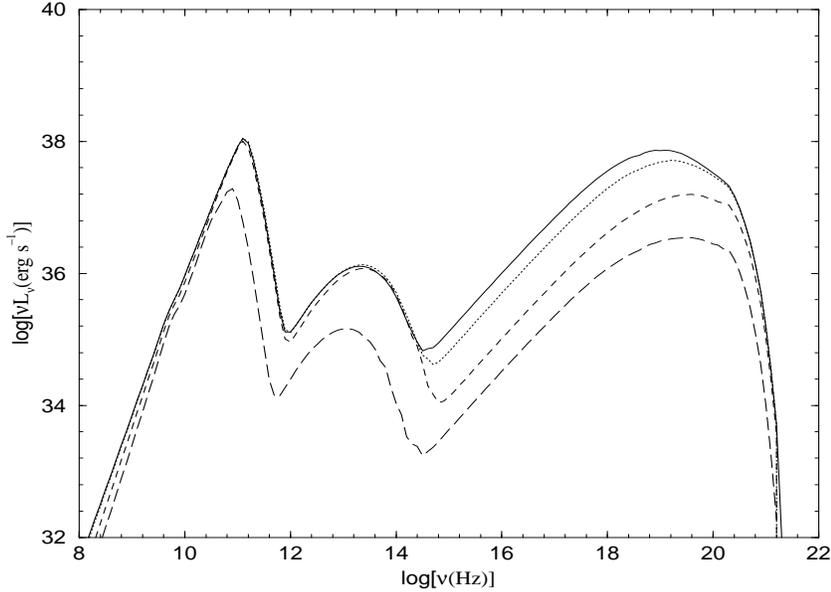, width=8cm, height=11.cm,angle=270}
\vskip -0.4cm
\caption{
The corresponding spectra of the solutions shown in Figure 2. Adopted
from Yuan et al. (2000).}
\label{fig3}\end{figure}

\section{An illustrative application to Sgr A$^*$}

As an illustrative example, we apply the above
results to the compact radio source Sgr A$^*$
located at the center of our Galaxy.
Advection-dominated accretion flow (ADAF) model has been turned out to
be of great success to explain its low luminosity and spectrum (Narayan, Yi \&
Mahadevan 1995; Narayan et al. 1998). However,
there exists a discrepancy between the mass accretion rate
favored by ADAF models in the literature
and that favored by the three dimensional hydrodynamical simulation,
with the former ($\sim 6.8 \times 10^{-5}\dot{M}_{\rm Edd}$,
see Quataert \& Narayan 1999) being
10-20 times smaller than the latter ($\sim 9\times 10^{-4} 
\dot{M}_{\rm Edd}$, see Coker \& Melia 1997). 
By seriously considering the
outer boundary condition of the accretion flow, we find that
due to the low specific angular momentum of the 
accretion gas (Coker \& Melia 1997), the
accretion in Sgr A$^*$ should belong to type III
which possesses a very large sonic radius.
This accretion pattern can significantly reduce the discrepancy between
the mass accretion rate, as Figure 4 shows (see Yuan et al. 2000 for details).

\begin{figure}
\vskip -0.4cm
\epsfig{file=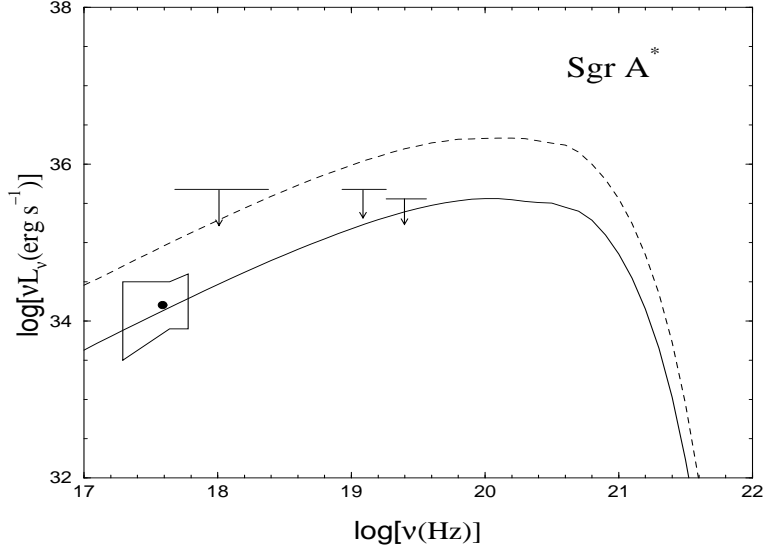, width=7.5cm, height=10.cm,angle=270}
\vskip -0.4cm
\caption{
The X-ray spectrum of Sgr A$^*$.
The observational
data are compiled by Narayan et al. (1998).
The spectra represented by the solid
and the dashed lines are produced by the accretion flows 
with the same accretion rate $\dot{M}=4 \times 10^{-4} \dot{M}_{\rm Edd}$
but different angular momentum at $r_{\rm out}$, 
$\Omega_{\rm out}=0.15 \Omega_{\rm K}$ 
for the solid line and
$\Omega_{\rm out}=0.46 \Omega_{rm K}$ for the dashed line.
Due to the difference of the angular momentum of the flow at the outer
boundary, the X-ray flux differs by a factor $\sim$ 8. 
Adopted from Yuan et al. (2000).}
\label{fig4}\end{figure}

\section{Discussion}

The present study is concentrated on the low-$\dot{M}$ case where
the differential terms in the equation such as the energy advection
play an important role therefore the effect of OBC are most obvious.
How about the role of OBC when  
the flows become optically thick?  
In this case, the electron and the ion possess
the identical temperature due to the strong couple between them and
the local viscous dissipation and radiation loss terms  in the energy balance
play an important role. As a result, the temperature profile 
is mainly determined {\em locally} rather than 
{\em globally} as in the case of optically thin flows.
Thus, the discrepancy of the
temperature caused by OBC will
lessen rapidly with the decreasing radii from the outer boundary.
This is also the reason why the temperature profiles
of one-temperature plasma and ions in Figures 1 and 2 converge
rapidly with decreasing radii. 
However, from our calculation to
the one-temperature accretion flow whose temperature
is also principally determined locally (Yuan 1999),
 we predict that the optically thick accretion 
flow should still present OBC-dependent
behavior in, e.g., the angular momentum and the Mach number
profiles which are in principle determined by the {\em momentum} 
rather than the {\em energy} equations. 
When the angular momentum of the accretion flow 
is less than a certain critical value, the accretion pattern should 
become of ``type III'' (Bondi-like). 
Although these conjectures need the confirmation
of detailed calculation, we note that the angular momentum profile
of slim disk model (see Figure 3 of  Abramowicz et al. 1988),
and a recent numerical simulation (Igumenshchev, Illarionov \&
Abramowicz 1999) seems to support this point.

Why the role of OBC in accretion disk models has been long neglected?
In the standard thin disk model, 
all the differential terms in the equations are neglected and the
differential equations are reduced into an algebraic one which 
don't entail any boundary conditions at all. 
In the later works on the global solutions 
for slim disks (Matsumoto et al. 1984; Abramowicz et al. 1988;
Chen \& Taam 1993) and optically thin advection-dominated
accretion flows  
(Narayan, Kato \& Honma 1997; Chen, Abramowicz \& Lasota 1997),
some authors did investigate the role of OBC, but failed to find
its importance. The main reason is that for optically
thick accretion flows (slim disk) or {\it one-temperature}
optically thin accretion flow, the local viscous dissipation 
plays an important role in the energy equation, so the effect of 
OBC lessen rapidly away from the outer boundary. In addition, the 
angular momentum in their outer boundary condition was always 
somewhat large. This might be the reason why they 
didn't find the solutions with
very large sonic radii.

\end{document}